\documentclass[fleqn,usenatbib]{mnras}

\usepackage{newtxtext,newtxmath}
\usepackage[export]{adjustbox}

\usepackage[T1]{fontenc}

\DeclareRobustCommand{\VAN}[3]{#2}
\let\VANthebibliography\thebibliography
\def\thebibliography{\DeclareRobustCommand{\VAN}[3]{##3}\VANthebibliography}

\usepackage{graphicx}	
\usepackage{amsmath}	
\usepackage{tabularx}   
\usepackage{dblfloatfix}      
\usepackage{xcolor}
\usepackage{upgreek}
\usepackage{textgreek}

\newcommand{\hii}{H\,{\sc ii}}

\title[Grain Depletions and Strong Spectra Lines]{Self consistent grain depletions and abundances II: Effects on strong-line diagnostics of extragalactic H II regions}

\author[C. M. Gunasekera et al.]{
Chamani M. Gunasekera,$^{1}$\thanks{E-mail:cmgunasekera@uky.edu ; xji243@uky.edu; mchatzikos@gmail.com; rbyan@cuhk.edu.hk; gary@g.uky.edu}
Xihan Ji,$^{1}$
Marios Chatzikos,$^{1}$
Renbin Yan$^{2}$
and Gary Ferland$^{1}$
\\
$^{1}$Department of Physics \& Astronomy, University of Kentucky, Lexington, KY 40506, USA\\
$^{2}$Department of Physics, The Chinese University of Hong Kong, Shatin, N.T., Hong Kong S.A.R., China\\
}

\date{Accepted 2023 January 20. Received 2023 January 11; in original form 2022 May 1}

\pubyear{2023}

\begin{document}
\label{firstpage}
\pagerange{\pageref{firstpage}--\pageref{lastpage}}
\maketitle

\begin{abstract}
The depletion of elements onto dust grains is characterized using a generalized depletion strength $F_*$ for any sightline, and trend-line parameters $A_X, B_X$ and $z_X$. 
The parameters $A_X, B_X$ and $z_X$ define the relative depletion pattern, for which values are published in previous works.
The present study uses these parameters to calculate post-depleted gas-phase abundances of 15 different elements while varying $F_*$ from 0 to 1. An analysis of emergent strong spectral line intensities, obtained by inputting the calculated abundances into a {\sc cloudy} model, shows that the depletion strength has a non-trivial effect on predicted emission lines and the thermal balance of the ionized cloud. The amount by which elements deplete also affects the coolant abundances in the gas. 
Furthermore, it was found that each of the parameters - metallicity, ionization parameter $U$ and depletion strength $F_*$ have degenerate effects on the emission-line strengths, and thermal balance of the interstellar medium (ISM). Finally, comparing our results to a sample of H\,{\sc ii} regions using data obtained from the 
Mapping Nearby Galaxies at Apache Point Observatory survey (MaNGA) revealed that the best-fit $F_*$ was approximately 0.5. 
However, this best-fit value does not work well for all metallicities.
Removing the sulfur depletion and changing the nitrogen abundance pattern can improve the fit. As a result, extra observational evidence is required to verify the choices of parameters and better constrain the typical depletion strength in galaxies.
\end{abstract}




\section{Introduction}
\label{sec:intro}
The condensation of various elements into solid form, or grain depletion, is an important physical process in the interstellar medium (ISM). Observations have long found that the elemental abundance in the ISM is lower than the values in stars \citep{1973ApJ...181L.122J,1973ApJ...181L.103M,1973ApJ...181L..97R,1973ApJ...181L.110R,1973ApJ...181L.116S,1978ppim.book.....S}. Despite the large number of observations, the physics of grain depletion is still not well understood. Observations of the absorption of the UV spectra of stars in the ISM of the Milky Way Galaxy have revealed that the degree of depletion varies between different heavy elements and at different locations of the ISM \citep{1973ApJ...181L.103M,1973ApJ...181L..97R,1973ApJ...181L.110R,1973ApJ...181L.122J}.
As a result, it is challenging to draw a consistent picture describing the depletion of all elements in the ISM. To solve the above problem, \cite{2009ApJ...700.1299J} (hereafter Jenkins09) studied the abundances along 243 different sightlines from more than 100 papers and proposed a unified scheme of parameterizing the depletion factors for a set of 17 different elements. The fact that the depletion factors of various elements are positively correlated makes it possible to describe them using a single parameter, which is denoted as $F_*$ by Jenkins09. The depletion factor ($F_*$) describes the overall dust depletion strength of a collection of elements, for a given line of sight. With the unification of the depletion patterns, it is now possible to estimate the typical level of depletion by using observations of a few lines.

Despite the important breakthrough by Jenkins09, this unified scheme for grain depletion has seldom been applied and studied in depth in works of theoretical photoionization modelling of the ionized ISM. This paper is the second of a series `{Self consistent grain depletions and abundances}'. The first paper \citep{2022MNRAS.512.2310G} (hereafter referred to as Paper\,{\sc i}) implements the Jenkins09 depletion model in {\sc cloudy} (a modelling software that can output spectral line predictions by simulating a broad range of conditions within the ISM \citep{2017RMxAA..53..385F}), to obtain photoionization model predictions of varying $F_*$ on the Orion Nebula as a test case. This work shows that values of depletion factors have a nontrivial effect on the model predictions for the Orion Nebula. On the one hand, grain depletion affects the remaining gas abundance of the depleted elements directly. On the other hand, it also impacts the thermal balance of the whole ionized cloud by changing the relative abundance of the coolants.
Inaccurate prescriptions of grain depletion would thus lead to unrealistic predictions on the emission-line spectra. As a consequence, the calibrations of various physical parameters in the ISM, including the gas-phase metallicity, would suffer from large uncertainties. Therefore, it is important to understand the effect of grain depletion on the theoretically predicted emission line spectra and the relevant emission line diagnostics. 

To better understand the effect of grain depletion on emission-line spectra and establish the results in Paper\,{\sc i}, the first part of this study extends the analysis used on the Orion Nebula to a more generalized photoionization model. Such an investigation utilizing a large range of $F_*$ values in conjunction with a range of ionization parameters and metallicities was made possible by the integration of the Jenkins09 model into {\sc cloudy} (see Paper\,{\sc i} for further description of this new function). The second part of this study compares these model predictions to observations of a sample of H\,{\sc ii} regions obtained from the Mapping Nearby Galaxies at Apache Point Observatory survey \citep[MaNGA,][]{bundy2015,yan2016b}.
The goal was to constrain the grain depletion factor for the general population of H\,{\sc ii} regions indirectly. Interestingly, even the best-fit depletion factor we found does not provide a perfect match to the data in a multidimensional line-ratio space. This is likely due to the N/O prescription we use to construct the models does not fit the data locus perfectly after we apply the depletion.

This paper is organized as follows. In Section~\ref{sec:method}, we describe the photoionization models we use and how we set-up the input parameters, especially the depletion factor. In Section~\ref{sec:results}, we present our results and show how the different line ratios depend on the depletion factor as well as its physical interpretation. In Section~\ref{sec:manga_data}, we discuss the effect of grain depletion on optical diagnostics of ionized regions and use observed emission-line spectra to constrain the depletion factor in extra-galactic H\,{\sc ii} regions. We conclude in Section~\ref{sec:conclusion}. For the frequently used logarithms of line ratios throughout this paper, log([O\,{\sc iii}] \textlambda5007/H$\beta$), log([N\,{\sc ii}] \textlambda6583/H$\alpha$), log([S\,{\sc ii}] \textlambda\textlambda6716,6731/H$\alpha$), and log([O\,{\sc i}] \textlambda6300/H$\alpha$), we denote them as R3, N2, S2,and O1, respectively.

\section{Photoionization models}
\label{sec:method}

The approach of this study is to use {\sc cloudy} to run model grids simulating H\,{\sc ii} regions by varying the metallicity (hereafter (O/H)/(O/H)$_{\odot}$, which is equivalent to (O/H)$_{\rm gas}$/(O/H)$_{\rm ref}$), ionization parameter, and the collective elemental depletion (represented by the depletion factor $F_*$) and predict the corresponding emission line spectra. This is the same procedure as described in Paper {\sc i}, but we generalized it for different ionization parameters and metallicities.

\begin{table}
	\caption{Reference abundance sets $(X_{\rm gas}/{\rm H})_{\odot}^{(2003)}$ \citep{lodders2003}, $(X_{\rm gas}/{\rm H})_{\odot}^{(2009)}$ \citep{2009M&PSA..72.5154L}, and GASS-\citep{2010Ap&SS.328..179G} are all available to be used with the {\sc cloudy} grain depletion command. The models for this study have used the GASS abundances as the reference set.}
	\label{tab:solar_abund}
	\begin{tabular}{lccc}	    
	\hline\hline
        Elem. & $(X_{\rm gas}/{\rm H})_{\odot}^{(2003)}$ & $(X_{\rm gas}/{\rm H})_{\odot}^{(2009)}$ & GASS \\
		\hline
        H	&	1.00E+00	&	1.00E+00  &   1.00E+00	\\
        He 	&	9.64E-02	&	8.43E-02  &   8.43E-02	\\
        Li	&	2.24E-09	&	1.90E-09  &   1.90E-09	\\
        Be	&	3.02E-11	&	2.09E-11  &   2.09E-11	\\
        B	&	7.08E-10	&	6.42E-10  &   6.42E-10	\\
        C  	&	2.88E-04	&	2.45E-04  &   2.45E-04	\\
        N	&	7.94E-05	&	7.24E-05  &   7.24E-05	\\
        O	&	5.75E-04	&	5.36E-04  &   5.36E-04	\\
        F	&	3.39E-08	&	2.74E-08  &   2.74E-08	\\
        Ne 	&	8.91E-05	&	1.12E-04  &   1.12E-04	\\
        Na	&	2.34E-06	&	1.97E-06  &   1.97E-06	\\
        Mg	&	4.17E-05	&	3.52E-05  &   3.52E-05	\\
        Al	&	3.47E-06	&	2.89E-06  &   2.89E-06	\\
        Si	&	4.07E-05	&	3.41E-05  &   3.41E-05	\\
        P	&	3.47E-07	&	2.83E-07  &   2.83E-07	\\
        S	&	1.82E-05	&	1.44E-05  &   1.44E-05	\\
        Cl	&	2.14E-07	&	1.76E-07  &   1.76E-07	\\
        Ar	&	4.17E-06	&	3.16E-06  &   3.16E-06	\\
        K	&	1.51E-07	&	1.28E-07  &   1.28E-07	\\
        Ca	&	2.57E-06	&	2.06E-06  &   2.06E-06	\\
        Sc	&	1.41E-09	&	1.17E-09  &   1.17E-09	\\
        Ti	&	1.00E-07	&	8.43E-08  &   8.43E-08	\\
        V	&	1.17E-08	&	9.76E-09  &   9.76E-09	\\
        Cr	&	5.25E-07	&	4.47E-07  &   4.47E-07	\\
        Mn	&	3.80E-07	&	3.15E-07  &   3.15E-07	\\
        Fe	&	3.47E-05	&	2.89E-05  &   2.89E-05	\\
        Co	&	9.55E-08	&	8.02E-08  &   8.02E-08	\\
        Ni	&	1.95E-06	&	1.67E-06  &   1.67E-06	\\
        Cu	&	2.19E-08	&	1.85E-08  &   1.85E-08	\\
        Zn	&	5.01E-08	&	4.44E-08  &   4.44E-08	\\
		\hline
	\end{tabular}
\end{table}

\subsection{The Jenkins09 depletion model} 
\label{sec:maths}
In the ISM, depletion of an element into dust grains is defined as the reduction of abundance below the expected abundance level. 
These expected abundances will be referred to as reference abundances henceforth, and denoted as $(X/{\rm H})_{\rm ref}$.
\begin{equation}
    [X_{\rm gas}/{\rm H}]_{F_*} = \log(X_{\rm gas}/{\rm H})_{F_*} - \log(X/{\rm H})_{\rm ref}.
\end{equation}

The study by Jenkins09 presents a linear relationship between the individual depletion of each element. They were able to find such a relation by introducing a generalized depletion strength $F_*$. This depletion strength describes the depletion of each element $[X_{\rm gas}/H]_{F_*}$ as a linear relation to itself ($F_*$), and it is common to all elements
\begin{equation}
    [X_{\rm gas}/{\rm H}]_{F_*}=B_X+A_X(F_*-z_X),
	\label{eq:jenkins2}
\end{equation}
where $A_X,B_X$ and $z_X$ (referred to as depletion parameters hereafter) are best-fit parameters found in table 4 of Jenkins09 for this linear relation. These depletion parameters are independent of the reference abundance because they were developed using differential changes in gas abundances as opposed to absolute depletions. Hence this study will use the \cite{2010Ap&SS.328..179G} reference abundances with Jenkins' depletion model, although his model was built using reference abundances obtained from \cite{lodders2003}.

\subsection{Input parameters for photoionization models}
\label{sec:model_param}

The photoionization models in this study is generated using the development version of {\sc cloudy} \citep[c17,][]{2017RMxAA..53..385F} (last described by \url{https://nublado.org/}). In addition to the overall chemical abundance and the dust prescription mentioned in the previous subsection, there are many other input parameters that impact the resulting emission line spectra. Here we summarize the values of the photoionization model parameters adopted in this work.

First, the ionization parameter ($U$) describes the relative strength of ionizing radiation in the H\,{\sc ii} region. $U$ is defined as $U\equiv \frac{\Phi _{\rm ion}}{n_{\rm H} c}$, where $\Phi _{\rm ion}$ is the flux of the hydrogen ionizing photons at the illuminated face of the cloud, $n_{\rm H}$ is the hydrogen density and $c$ is the speed of light. This study varies the ionization parameter in the logarithmic space so that $-4.0 < \log (U) < -2.0$ with increments of $0.5$. The shape of the radiation field is described using the stellar SEDs generated by the code {\sc starburst99} \citep[v7.01,][]{leitherer1999,leitherer2014}. We adopt a Kroupa IMF \citep{kroupa2001}, a continuous star formation history of 4 Myr, the Geneva evolutionary track with standard mass-loss rates, and the Pauldrach/Hiller model atmosphere \citep{pauldrach2001, 1998ApJ...496..407H}. A total of six SEDs with stellar metallicities from $\log (Z/Z_\odot )=-1.3$ to $\log (Z/Z_\odot )=0.5$ are considered. Since {\sc starburst99} only provides models with stellar metallicity up to $\log (Z/Z_\odot )\approx 0.3$, we compute the SED model with the highest stellar metallicity through linearly extrapolating the logarithmic fluxes. When these SEDs are used in our photoionization models, the gas-phase metallicity, [O$\rm _{\rm gas}$/H], is ensured to match the stellar metallicity.

For the structure of the ionized cloud, a plane-parallel geometry was chosen and calculations are stopped when the temperature falls below 100 K. The equation of state is set to be isobaric and the initial hydrogen density is 14 cm$^{-3}$, which is the median density found in the SF regions of MaNGA \citep{ji2020a}. A detailed description of the MaNGA survey is delayed to Section~\ref{sec:manga_data}.

To fix the collective depletion of the elements, we use the unified depletion model provided by Jenkins09 described in Section~\ref{sec:maths}. Paper {\sc i} has introduced a command that implements this Jenkins09 model into {\sc cloudy}, which streamlines the calculation of depleted gas-phase elemental abundances and allows for a grid of varying $F_*$ values to be simulated.
Since the equations and parameters used to calculate the input abundances are from the Jenkins09 study, the complete range of depletion factors used in this study is from $F_* = 0$ to $1.0$. The present study breaks this $F_*$ range into increments of $0.125$. It should be noted that these upper and lower limits of $F_*$ are not real physical limits; they are the highest and lowest observed depletions from the observations compiled by Jenkins09. As such, $F_* = 0$ should be interpreted as the system having some small amount of depletion above no depletion.

The abundance of dust grains is scaled with $F_*$ as described in Paper {\sc i}. The default set of grain abundance used by {\sc cloudy} is assumed to correspond to a depletion strength of $F_* = 0.5$, as before. The subsequent grain abundances for varying $F_*$ are scaled with the fraction of total $(X_{\rm grains}/{\rm H})$ at a given $F_*$ to the total $(X_{\rm grains}/{\rm H})$ at $F_*=0.5$, as calculated using the Jenkins09 depletion function. Since any element that depletes from the gas phase should appear in the dust phase, we define grain abundance as, 
\begin{equation}
    (X_{\rm grains}/{\rm H}) \equiv 1 - (X_{\rm gas}/{\rm H}).
\end{equation}

The metallicity ((O/H)/(O/H)$_{\odot}$) of the model input into {\sc cloudy} was altered using the ‘metallicity scale factor’ command. {\sc cloudy} multiplies the abundances given in the abundance file, by the value provided in this command line. Note that using (O/H)/(O/H)$_{\odot} = 1.00$ means that the model has gas abundances equivalent to the reference abundances. Although the Jenkins09 model was built using \citep{lodders2003} and a newer set of abundances are available from the same authors -- \citep{2009M&PSA..72.5154L}, we use the reference abundance set of GASS. All three reference abundance sets are summarized in Table~\ref{tab:solar_abund}. (O/H)/(O/H)$_{\odot}$ used in this study are $0.05$ which represents a metal-poor system, $1.00$ which keeps the reference abundance set unchanged, and $3.16$ which represents a metal high set. The value of the factor for the highest metallicity model, $3.16$, is derived from extrapolating the SED of starburst99 in order to cover the whole data space used in Section~\ref{sec:manga_data}. Models with metallicity scale factors $0.20, 0.40$ and $2.00$ were also studied, but are only shown and discussed in Section~\ref{sec:manga_data} for the sake of brevity.

Finally, the abundances of some specific elements require additional treatments. For helium, we consider the cosmic plus nuclear synthesis production of these elements using the formula of \cite{2002ApJ...572..753D}
\begin{equation}
    \rm He/H = 0.0737 + 0.024 (Z/Z_\odot ).
\end{equation}
For secondary elements including carbon and nitrogen, we use the prescription of \cite{2013ApJS..208...10D} while refitting their relation using a second-order polynomial function
\begin{equation}
    \rm N/O = 0.0096 + 72\cdot \text{O/H} + 1.46\times 10^4 \cdot \text{(O/H)}^2,
\end{equation}
which gives the N/O ratio at each metallicity. In addition, we fixed C/N to be the reference abundance value, as did \cite{2013ApJS..208...10D}. We note that this relation only describes the nitrogen-to-oxygen ratio before depletion.

A summary of all the model parameters can be found in Table.~\ref{tab:model_param}.
\begin{table}
	\centering
	\caption{Photoionization model parameters}
	{\fontsize{8}{9.5}\selectfont
	\label{tab:model_param}
	\begin{tabularx}{\columnwidth}{ll}
	    \hline\hline
        Parameter & Input value\\
        \hline
        F$_*$ & 0, 0.125, 0.25, 0.375, 0.5, 0.625, 0.75, 0.875, 1\\
        Z/Z$_\odot$ & 0.05, (0.20, 0.40), 1.00, (2.00), 3.16\\
        log(U) & $-$4.0, $-$3.5, $-$3.0, $-$2.5, $-$2.0\\
        SED & Continuous star formation over 4 Myr\\
         &  (generated by {\sc starburst99})\\
         Secondary element & Same as the ones adopted by \cite{2013ApJS..208...10D}\\ 
        prescriptions & \\
        Background radiation & Cosmic ray background at $z = 0$\\
        Geometry & Plane-parallel\\
        Initial hydrogen density & $\rm 14~ cm^{-3}$\\
        Equation of state & Constant gas pressure\\
	    \hline
	\end{tabularx}
	}
\end{table}

\section{Impact on a generalized H\,{\sc ii} region}
\label{sec:results}
The collisionally-excited lines we study in this work are produced by collisions between heavy element atoms and free electrons that result in the excitation of an atomic electron, followed by its radiative de-excitation.
Various factors can alter the rate of these collisions, such as increasing temperature which increases the rate of collisions, and changing the abundance of atoms per H available for collisions. Since variations in the depletion strength, $F_*$, within a gas cloud alters the abundance pattern, we expect that $F_*$ affects both of the above-mentioned factors. An analysis of how the gas temperature and the emission line spectra behave with variations in $F_*$ in conjunction with variations in ionization parameter and metallicity should provide insight into how these parameters may be constrained in photoionization models. 

The calculation of post-depletion abundance of each element has been streamlined by incorporating the Jenkins09 depletion model and their depletion parameters into the widely used program {\sc cloudy} as described by Paper {\sc i}. That paper provides a preliminary study into how the depletion factor $F_*$ affects the spectra of a benchmark H\,{\sc ii} region -- the Orion Nebula. In the present study, we shall extend this work to a more generalized H\,{\sc ii} region. In doing so, we have run the photoionization model described in Section~\ref{sec:model_param} with three variable dimensions -- $F_*$, $U$, and metallicities.
The resulting spectral line intensities of R3, N2, S2, and O1, as well as the electron temperature profiles obtained are discussed in the following sections. 

\subsection{Effect of depletion strength on temperature}
\label{sec:fstar_Temp}
\begin{figure*}
    \includegraphics[width=\textwidth]{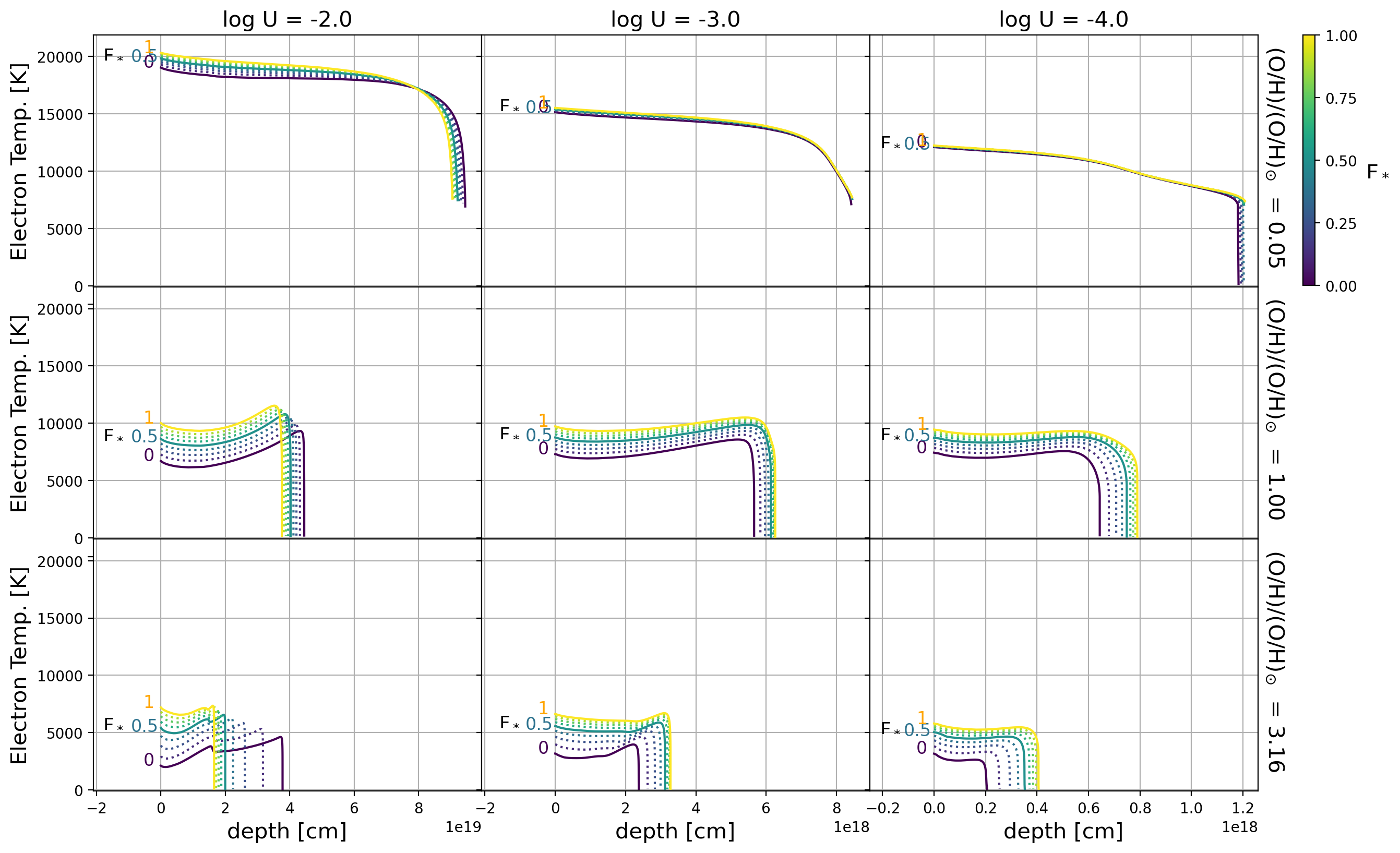}
    \caption{Temperature profile of the cloud at varying metallicity values (listed on the right edge of the figure), ionization  parameters (listed on the top edge of the figure), and depletion strength. Each line jumps by $0.125$ change in $F_*$. The depth is taken from the illuminated face at 0 cm, increasing out to the H ionization front. The ionization front is seen where the temperature profile drops off. In the bottom-left most panel, the second bump in the profile indicates the He ionization front. Despite the x-axis scale, this figure can be related to the Str\"omgren length via the ionization parameter, see Equation~(\ref{eq:strom_L}).}
    \label{fig:temp_F*}
\end{figure*}

The equilibrium temperature of an interstellar cloud in (or close to) thermal balance should remain relatively constant over time, i.e. the cooling mechanisms should balance the heating mechanisms of the gas. In H\,{\sc ii} regions photoionization of H atoms dominate in heating the gas, while inelastic collisions between electrons and ions dominate in cooling the gas. 
Sulfur, oxygen and nitrogen are important coolants in the H\,{\sc ii} region. 
Therefore, we expect temperature changes to be dominated by changes in the gas-phase abundances, particularly of S and O. 
In Paper {\sc i}, we showed that changing $F_*$ had two effects on the temperature profile of the Orion model. Higher depletion strengths both increase the overall temperature of the Orion model and cause the H ionization front to occur at shallower depths.
Since the depletion strength affects the abundance of gas-phase elements, which includes the abundance of the above-mentioned coolants, we expect a similar trend for the temperature profile in this model to that shown in Paper {\sc i}. 

Fig~\ref{fig:temp_F*} presents the electron temperatures, $T_{\rm e}$, as a function of depth of \hii\ regions with varying $U$, (O/H)/(O/H)$_{\odot}$, and $F_*$.
The electron temperature $T_{\rm e}$, which is the kinetic temperature of charged particles, provides a measure of the \hii\ region temperature. We vary two parameters in this study which were kept constant in Paper {\sc i} -- metallicity and ionization parameter, to understand how other parameters may affect the \hii\ region in conjunction with $F_*$.
A further discussion of why we expect temperature changes to be dominated by changes in the gas-phase abundances, particularly of S and O, is included in Section~\ref{sec:fstar_LineStrngth}.

For all metallicity cases, we find that increasing $F_*$ or $U$ prompts higher electron temperatures. From each panel of Fig~\ref{fig:temp_F*}, we see that the overall electron temperature rises with depletion strength (the purple lines are at the lowest temperature values, and the temperature gradually climbs up to the yellow lines per panel). This is due to the fact that $F_*$ increasingly depletes the gas-phase abundances of coolants. 
Moreover, we find that the correlation between electron temperature and ionization parameter is a result of more H photoionizing interactions that heats the gas with higher values of $U$.
In contrast, increasing (O/H)/(O/H)$_{\odot}$ cools down the system, as this amplifies the abundance of coolants (compare plotlines of each colour between the three panels in Fig~\ref{fig:temp_F*}). 

A higher (O/H)/(O/H)$_{\odot}$ gives rise to a greater variation in temperature as a function of $F_*$.
This effect is observed in Fig~\ref{fig:temp_F*}, where the plot lines span a bigger range of temperatures moving down the panels. 
Increasing (O/H)/(O/H)$_{\odot}$ results in a larger fraction of element abundances being condensed into dust grains i.e. more heating by grains. 
As such we find that metallicity makes the temperature of the \hii\ region more sensitive to changes in $F_*$.

For a given (O/H)/(O/H)$_{\odot}$, temperature changes become increasingly smaller with $F_*$. In each panel of Fig~\ref{fig:temp_F*}, when comparing the spacing of the plot lines between the different values of $F_*$, the spacing becomes smaller with $F_*$. This is a direct result of the depleted fraction of an element $X$ being modeled by an exponentially decreasing function, $10^{B_X + A_X(F_*-z_X)}$ where $A_X, B_X < 0$. Hence for larger $F_*$ values, the temperature of the \hii\ region becomes less sensitive to the depletion strength.

The final observed effect is that the location of the hydrogen ionization front is affected by all three variables $F_*$, (O/H)/(O/H)$_{\odot}$, and $U$. The H ionization front is the boundary where the neutral ISM (a.k.a H\,{\sc i} region) meets the edge of the H\,{\sc ii} region, and so its location is affected by the abundance of photons available to ionize the gas. In addition, the Str\"omgren length $L$ of an ionized layer is related to the ionization parameter via the following equation,
\begin{equation}
    L = \frac{Uc}{n_p\alpha_{\rm B}(T_{\rm e})},
    \label{eq:strom_L}
\end{equation}
where,
\begin{equation}
    \alpha_{\rm B}(T) = \begin{cases} 2.90\times10^{-10} T_{\rm e}^{-0.77}, & T_{\rm e} \leq 2.6\times10^4 {\rm K} \\ 1.31\times10^{-8} T_{\rm e}^{-1.13}, & T_{\rm e} > 2.6\times10^4 {\rm K} \end{cases},
    \label{eq:alpha}
\end{equation}
\citep{1998PASP..110.1040B, 1980PASP...92..596F}.
First of all, our study has confirmed that raising $U$ results in a larger ionized layer, as can be observed from the H front at continuously larger depths going from the right to the left of Fig~\ref{fig:temp_F*}. This is understood by the fact that greater ionization parameters expand the H$^+$ layer by ionizing more H atoms.
Secondly, we can see from Fig~\ref{fig:temp_F*}, the electron temperatures of our model \hii\ region ranges from $~2500$ K to $~20$ $000$ K. So, for a constant ionization parameter (in the range $10^{-4}\leq U <10^{-2}$), any positive changes in temperature result in a lengthened $L$ according to equations~(\ref{eq:strom_L}) \&~(\ref{eq:alpha}) (a result of an increase in the rate of H ionization collisions). This is exactly the result observed in the center and rightmost panels of Fig~\ref{fig:temp_F*}. In contrast, our figure shows the opposite trend at $U = 10^{-2}$. At the large $U$ limit of over $10^{-2}$, grains dominantly absorb the ionizing photons, instead of hydrogen \citep{1998PASP..110.1040B}. Since a larger depletion strength results in a larger dust abundance, a greater fraction of ionizing photons are absorbed by those dust grains, leading to the Str\"omgren sphere shrinking. This is the trend reflected in the leftmost panels of Fig~\ref{fig:temp_F*}, where we observe the H ionization front occurring at shallower depths with increasing $F_*$. This is also the trend observed in Paper {\sc i}, in which $\log{U} = - 1.48$. Lastly, since metallicity increases the abundance of dust grains and reduces the temperature, the H ionization front occurs at shallower depths, for any given $U$ and $F_*$.

We conclude that electron temperature is anticorrelated to metallicity, while directly correlated to the depletion strength and ionization parameter. In addition to the $F_*$ to electron temperature relation already established in Paper {\sc i}, here we include the relations to two additional parameters that are fundamental to \hii\ regions.
We shall see in the next two sections (Section~\ref{sec:fstar_U} and Section~\ref{sec:fstar_LineStrngth}) that these results follow from changes to coolant abundances, with (O/H)/(O/H)$_{\odot}$ and $F_*$ working in tandem. Furthermore, the dependency of the electron temperature on $F_*$ seems to change with $F_*$ itself and (O/H)/(O/H)$_{\odot}$. Moreover, we shall see that the sum of coolant line strengths is affected by depletion strength in a similar trend to how electron temperature is affected by $F_*$. Finally, we have discovered two regimes of how $F_*$ affects the location of the H ionization front based on the ionization parameter. In Paper {\sc i}, we had already observed the high ionization parameter behaviour -- the depth of the H front being anticorrelated to $F_*$. Here we find the second regime, where at low ionization parameters, the depth of this ionization front is positively correlated to $F_*$.

\subsection{Effect of depletion strength on emission line intensities}
\label{sec:fstar_LineStrngth}
\begin{figure}
    \includegraphics[width=\columnwidth]{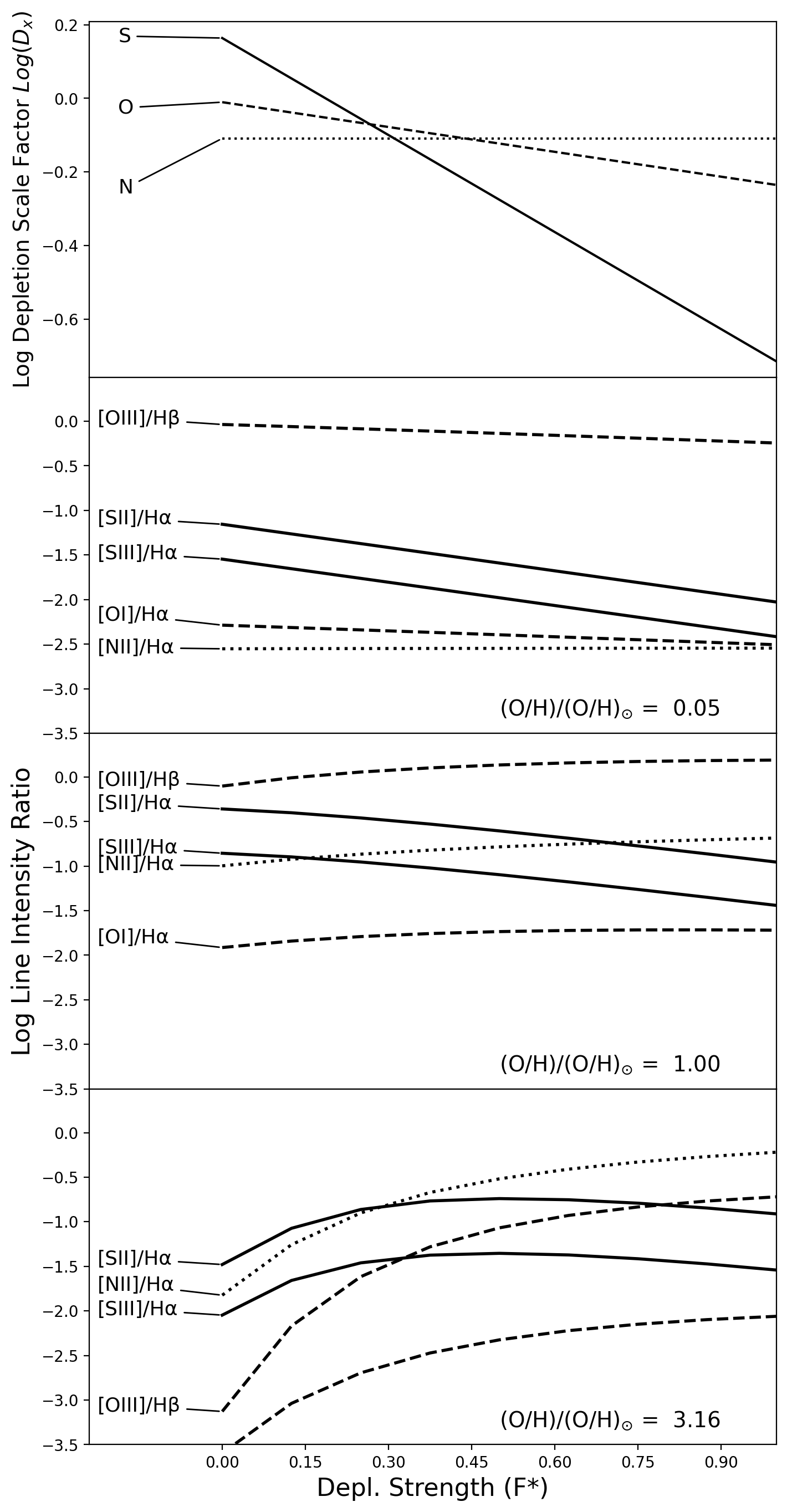}
    \caption{Top panel: the depletion scale factor calculated from the Jenkins09 depletion model, and multiplies the reference abundance in logarithm scale as a function of $F_*$. The depletion scale factor $D_X$ is defined using $(X_{\rm gas}/{\rm H})_{F_*} \equiv (X/{\rm H})_\odot D_X$. In the case of the Jenkins09 depletion model we have that, $\log(D_X) = B_X+A_X(F_*-z_X)$. Bottom three panels: line ratio intensities obtained from the {\sc cloudy} model predictions as a function of $F_*$, for the three different metallicity values indicated. These line ratios correspond to fixed ionization parameter at $\log (U) = -3.0$.}
    \label{fig:line_strength}
\end{figure}

The heavy element atoms in the gaseous state of \hii\ regions collide with free electrons, and the atomic electrons become excited. An emergent emission line spectrum from the gas results from the de-excitation of these electrons. 
Since the depletion strength affects the abundance of atoms available for these collisions, and the temperature of the gas which in turn affects the rate of these collisions, we expect that changing $F_*$ will affect the emission spectrum emergent from the \hii\ region. 
Evidence of this behaviour was presented in Paper {\sc i}, where we found two competing effects in the Orion model. We found that increasing the depletion of heavy elements, weakened the corresponding line intensities i.e. the intensity of S2; increasing the abundance of grains, while depleting coolants, increased the temperature which thereby enhanced the collisionally excited line ratios corresponding to minimally depleted elements i.e. the intensities of R3, N2, and O1. Here, we include a fourth emission line ratio log([S\,{\sc iii}] \textlambda9530/H$\alpha$) to confirm the behaviour of emission lines corresponding to ions of sulfur.

In comparing the line ratios of the different ions in Fig~\ref{fig:line_strength}, we observe that the S2 and S3 ratios (which are related by a factor, for all depletion strengths) exhibit different trends to the ratios of other ions. Similar trends were observed for Orion, in Paper\,{\sc i}. For (O/H)/(O/H)$_{\odot}<1$, these two lines are more strongly anticorrelated to $F_*$, than the other lines are. In the previous section, it was found that $F_*$ has little effect on temperature, as seen in Fig~\ref{fig:temp_F*}. This is indicative of line ratio trends being predominantly a result of the selective depletions of heavy elements at low metallicities. Hence the emission line ratios corresponding to heavily depleted elements reduce with $F_*$.

At higher metallicities than the reference abundances, N2, O3, and O1 are more strongly correlated with $F_*$ in comparison to S2. Additionally, there is a transition from a linear relation with $F_*$ to a non-linear one. Recall from Section~\ref{sec:fstar_Temp} at high (O/H)/(O/H)$_{\odot}$, we observed a significant change in the temperature of the gas with $F_*$, and that they have an non-linear relation. 
As the line ratios are seen to reflect the same trend with $F_*$ as the temperature does, at (O/H)/(O/H)$_{\odot}>1$, we find that the temperature of the gas dominantly affects the line ratios.
This is evident in the fact that according to the Jenkins09 depletion pattern, $F_*$ does not influence the depletion of nitrogen. However, at the reference and higher metallicities, the N2 line ratio intensifies with an increase in $F_*$ as a result of a higher rate of collisional de-excitations with temperature.

Finally for the highest metallicity case and $\log(U)=-3$, we find that there is a shift in the dominant coolant as sulfur becomes heavily depleted at high $F_*$. At (O/H)/(O/H)$_{\odot}=3.16$ and $F_*=0$ our model outputs tell us that sulfur is responsible for $19.8$ per cent of the total cooling while oxygen and nitrogen is responsible for $<1$ per cent cooling each (the top panel of Fig~\ref{fig:line_strength} shows that here, there is little to no depletion in S compared to O and N). Increasing $F_*$ to $1$, increases the fraction of total cooling by both oxygen and nitrogen to $17.5$ per cent each, while sulfur reduces to $5.5$ per cent of the cooling (the top panel of Fig~\ref{fig:line_strength} shows that, here S is much more heavily depleted than O and N). This is apparent in the bottom panel of Fig~\ref{fig:line_strength}. At high (O/H)/(O/H)$_{\odot}$ and high $F_*$, while all other line ratios increase with $F_*$, both S2 and S3 decrease. As the temperature plateaus with increasing $F_*$, the total rate of cooling should also plateau with $F_*$. However, at this level of depletion, sulfur is heavily depleted, making oxygen and other less depleted coolants dominate the cooling mechanisms. 

In conclusion, there are two competing effects on the emission line spectra as a function of depletion strength, based on the metallicity. At low metallicities, the dominant effect on the line intensities is caused by the change in abundance pattern with $F_*$. This was the regime observed in Paper {\sc i} in the Orion Nebula. While at high metallicities the dominant effect on the line intensities is caused by variations in temperature with $F_*$. Finally, there is a shift in oxygen and nitrogen becoming one of the dominant coolants as $F_*$ increases and sulfur along of with other heavily depleted coolants become unavailable to maintain the rate of cooling.

\subsection{Effect of depletion strength \& ionization parameter on BPT-like diagnostics}
\label{sec:fstar_U}

\begin{figure*}
    \includegraphics[width=0.97\textwidth, left]{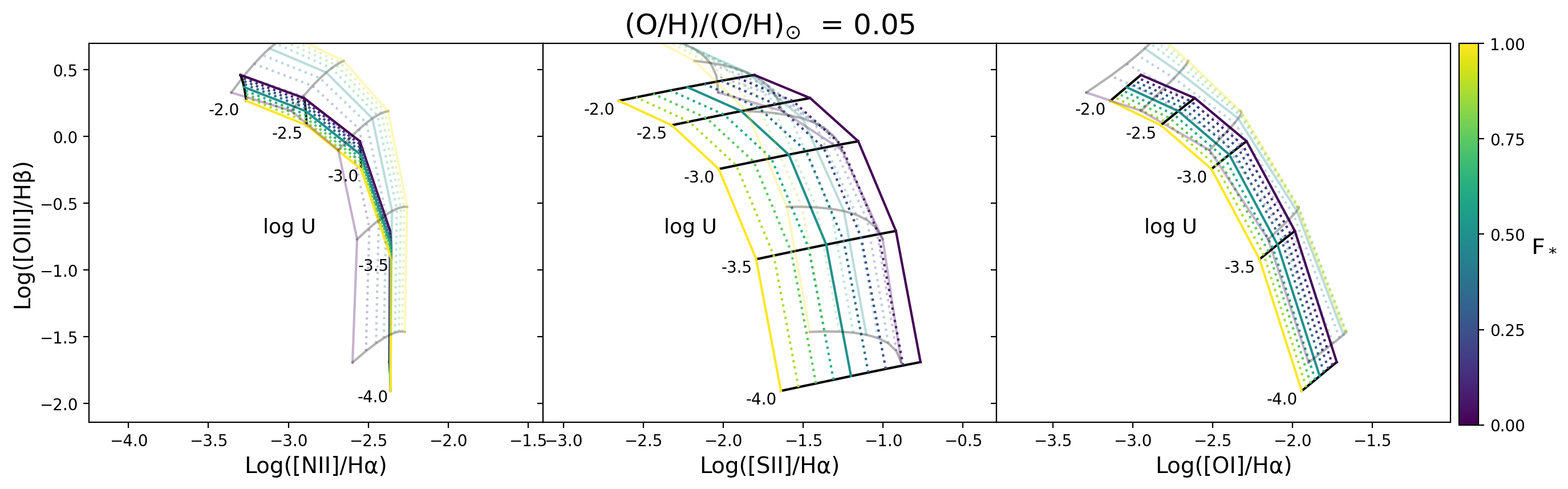}
    \includegraphics[width=0.91\textwidth, left]{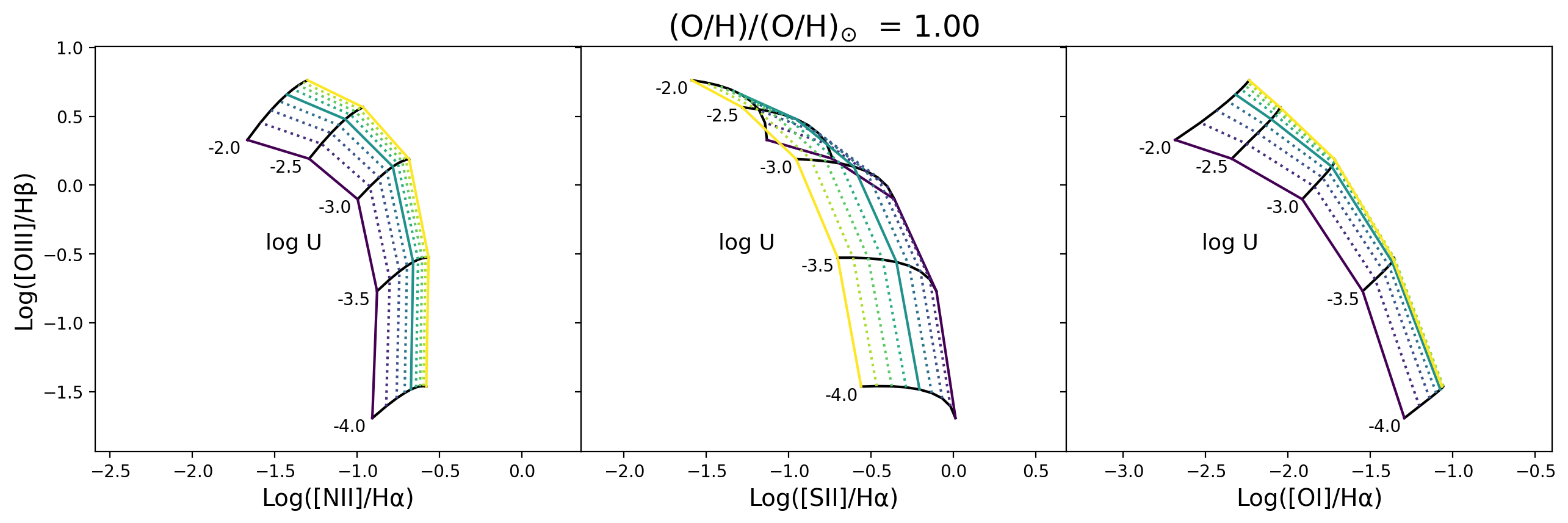}
    \includegraphics[width=0.91\textwidth, left]{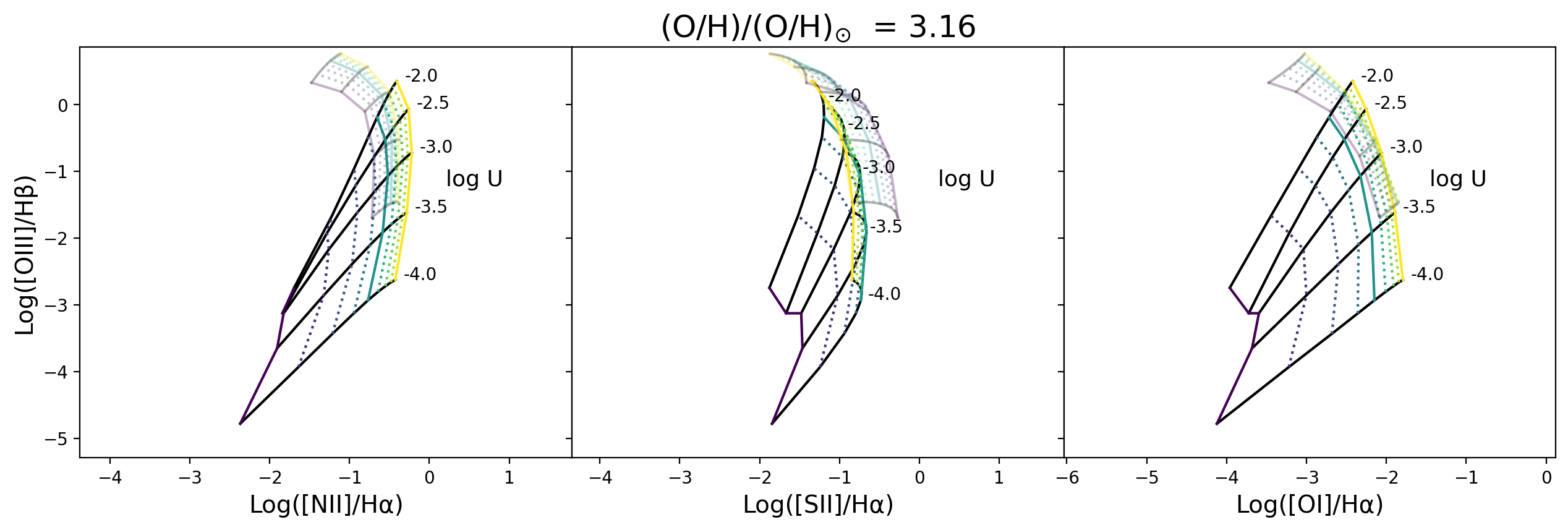}
    \caption{BPT-like plots using the spectral line ratio predictions for models of varying (O/H)/(O/H)$_{\odot}$ (top to bottom rows of plots), varying depletion strength $F_*$ (coloured lines), and varying U (black lines) in [N\,{\sc ii}]-, [S\,{\sc ii}]-, and [O\,{\sc i}]-BPT diagrams. The line ratios for varying $F_*$ are the same ones depicted in Fig~\ref{fig:line_strength}, and the colour coding for models of varying $F_*$ is the same as that in Fig~\ref{fig:temp_F*}. Ghost plots of the reference metallicity (middle row) plots are included in the background of the highest (bottom row) and lowest (top row) metallicity plots to facilitate comparisons.}
    \label{fig:bpt}
\end{figure*}

The ionization parameter, being the ratio of ionizing photon flux to gas density, directly affects the intensity of H recombination lines. In contrast, depletion strength directly affects the grain abundance in the \hii\ region. Since Hydrogen and dust grains affect the thermal balance of the gas (as seen in Section~\ref{sec:fstar_Temp}), we expect $U$ and $F_*$ to exhibit competing effects on the emergent line intensities, thereby affecting BPT diagnostic plots.
In Paper {\sc i}, we found that increasing $F_*$ increased the temperature and thus strengthened the collisionally excited lines of minimally depleted elements in the Orion Nebula. Here we study the effects of varying an additional parameter $U$ along with $F_*$ on a generalized \hii\ region.

Optical diagnostic diagrams \citep[BPT diagrams hereafter,][]{1981PASP...93....5B, 1987ApJS...63..295V, ho1997} composed of pairs of strong optical emission line ratios are very powerful in distinguishing different ionized regions in galaxies (H\,{\sc ii} regions, active galactic nuclei, low-ionization nuclear emission line regions, etc.).
In order to understand how dust depletion affects this diagnostic diagram in Fig~\ref{fig:bpt}, we present BPT plots of our models with varying $U$ and $F_*$ in the [N\,{\sc ii}]-, [S\,{\sc ii}]-, and [O\,{\sc i}]-BPT diagrams. These graphs are constructed for the three metallicities listed as before, and the colour coding for $F_*$ remains the same as that in Fig~\ref{fig:temp_F*}. The results for fixed $U$ are represented by the black plot lines, and those for fixed $F_*$ are represented by the coloured plot lines.

Fig~\ref{fig:bpt} shows that, at all three metallicities, both $U$ and $F_*$ work in tandem to affect the spectral line ratios.
We find that $U$ is more strongly correlated to R3 than the other line ratios presented in this study (the black plot lines span a wider range of R3 intensities than the other line intensities). This is a result of R3 having a much higher ionizing potential than N2, S2,or O1. At low metallicities, since $F_*$ has little effect on R3, we find that varying $F_*$ does little to change the behaviour of the line ratios as a function of $U$. However, at high metallicity, since both $U$ and $F_*$ are most strongly correlated to R3, the shape of the emission ratio, $U$ plot changes non-trivially with the value of $F_*$. In this regime, increasing $F_*$ enhances the effects of $U$ on R3, while diminishing its effects on the other emission ratios.

A second prominent feature of this figure is that the (O/H)/(O/H)$_{\odot}$ affects whether $F_*$ and $U$ are positively or negatively (anti-) correlated with the emission line ratios. $U$ is anticorrelated to N2, S2,and O1 line strengths at the lowest (O/H)/(O/H)$_{\odot}$, while $F_*$ is anticorrelated with only S2 and O1. There is no change in N2 with $F*$ and R3 increases with both $U$ and $F$.
In contrast at the highest (O/H)/(O/H)$_{\odot}$, both $F_*$ and $U$ are positively correlated with the line intensity ratios.
Increasing the overall degree of ionization in the gas cloud results in more ionized atoms available for collisions with free electrons intensifying the emission lines. At high (O/H)/(O/H)$_{\odot}$, the amount in which the low and high-ionization lines are intensified relative to the Balmer lines is greater.
Finally, at the reference (O/H)/(O/H)$_{\odot}$, N2, and O1 are directly correlated with $F_*$, and S2 is anticorrelated with $F_*$ -- a combination of the effects we see at the high and low (O/H)/(O/H)$_{\odot}$ extremes. 

In conclusion, we have found that all three parameters -- $F_*$, $U$, and (O/H)/(O/H)$_{\odot}$ affect how variations in the other two parameters change the shape of the BPT-diagnostic plots. Thus, in order to constrain $F_*$ in an ionized region, we must also constrain both metallicity and ionization parameter. In Section~\ref{sec:manga_data}, we provide a discussion on constraining $F_*$ for a set of \hii\ regions obtained from the MaNGA survey, by comparing the observed data to our photoionization model predictions.

\section{Impact of varying $F_*$ on optical emission-line diagnostics of observed ionized regions}
\label{sec:manga_data}

Photoionization models set the theoretical basis for defining the demarcation lines in BPT diagrams 
\citep[e.g.][]{kewley2001, kewley2006}, and can be used to calibrate the physical properties of ionized clouds based on the observed ratios of strong emission lines \citep[see][and references therein]{kewley2019}. The intrinsic shapes of the photoionization models in different emission-line ratio spaces also offer a possibility to define new diagnostic diagrams that orient the models in desired positions, thus facilitating constraints on the model parameters \citep{vogt2014, 2020MNRAS.499.5749J}.

Despite the great success of the BPT diagrams and their variants, the details of the built-in assumptions, including the intrinsic variations in ionizing sources and ionized regions in the same class, are seldom examined carefully. As we have seen in previous sections, variations in the depletion factor $F_*$ result in a non-trivial change in the observed line ratio, depending on the species of ions involved. Quite often a single set of depletion factors is assumed in practices of comparing photoionization models with observational data. This effectively enforces a single $F_*$ value for the entire sample. Therefore, it is important to understand whether a single $F_*$ suffices to describe the general population of a class of ionized regions (e.g. H\,{\sc ii} regions).

\begin{figure*}
\includegraphics[width=1.0\textwidth]{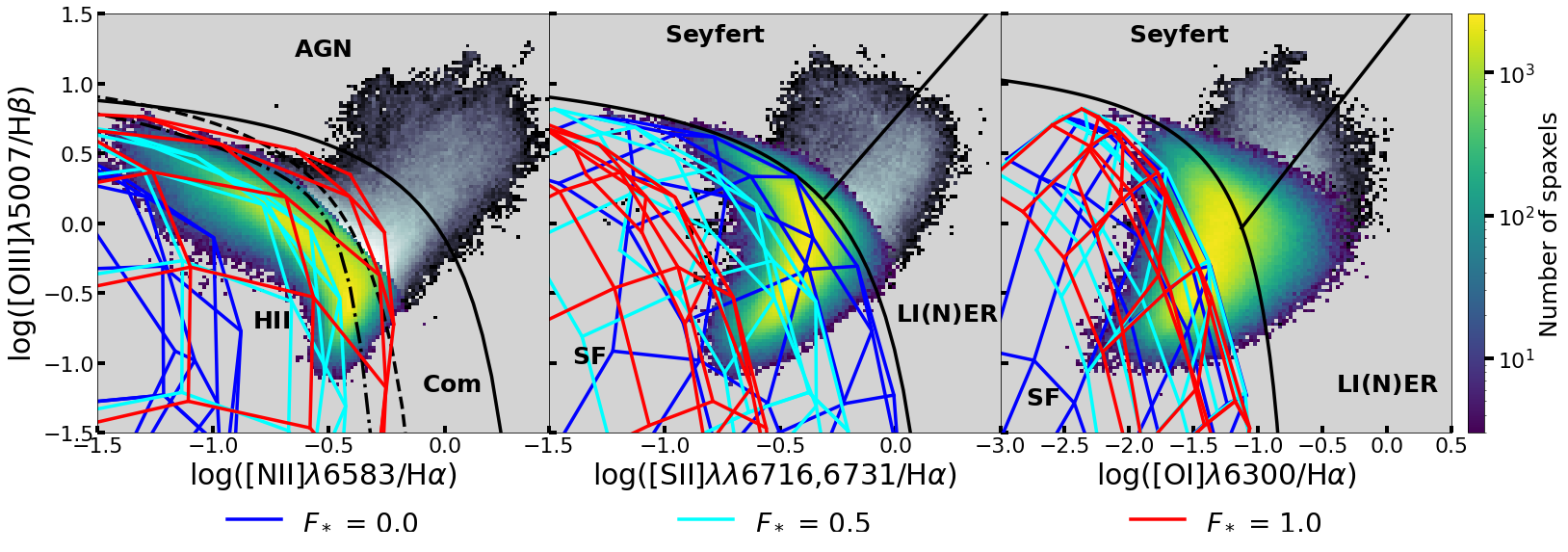}
\caption{Photoionization models with $F_*$ = 0.0, 0.5, and 1.0 plotted in the [N\,{\sc ii}]-, [S\,{\sc ii}]-, and [O\,{\sc i}]-BPT diagrams. The model grids are plotted by independently varying the metallicity and the ionization parameter. A sample of spatially resolved data from MaNGA is shown in the background, with the SF regions (identified by combining the [N\,{\sc ii}]- and [S\,{\sc ii}]-BPT diagrams) coloured from yellow to green, and the rest of the ionized regions coloured from white to black.
We also plot demarcation lines from \protect\cite{kewley2001} (solid black lines), \protect\cite{kewley2006} (solid black lines), \protect\cite{kauffmann2003} (dashed black line), and \protect\cite{stasinska2006} (dash-dotted black line) for comparison.
}
\label{fig:3bpt}
\end{figure*}

In this section, we aim to select a sample of observed H\,{\sc ii} regions and constrain $F_*$ by comparing them with the {\sc cloudy} models we computed in this work. Our data are drawn from SDSS-IV MaNGA \citep{bundy2015,yan2016b}, which includes a great number of H\,{\sc ii} regions to large effective radii. The primary sample galaxies in MaNGA are observed out to at least 1.5 $\rm R_e$, and the secondary sample galaxies are observed out to at least 2.5 $\rm R_e$ \citep{wake2017}. We use the public data of MaNGA, which is part of the $\rm 15^{th}$ data release of SDSS \citep[DR15,][]{aguado2019}. Fig.~\ref{fig:3bpt} shows the distribution of MaNGA's spatial pixels (spaxels) in the three most widely used BPT diagrams. For these sample spaxels, the signal-to-noise ratios (S/N) of H$\alpha$, H$\beta$, [S\,{\sc ii}] \textlambda\textlambda$6716, 6731$, [N\,{\sc ii}] \textlambda$6583$, [O\,{\sc i}] \textlambda$6300$, and [O\,{\sc iii}] \textlambda$5007$ are all required to be greater than 3. One can see that the original extreme starburst lines proposed by \cite{kewley2001} are not very good representations of the upper envelopes of H\,{\sc ii} or star-forming (SF) regions in these diagrams. As noted by \cite{belfiore2016} and \cite{law2021}, H\,{\sc ii} regions residing in large radial positions in galaxies can cross the extreme starburst lines in the [S\,{\sc ii}]- and [O\,{\sc i}]-BPT diagrams. In light of this, we select a sample of H\,{\sc ii} regions not based on the original BPT diagrams, but using the three-dimensional diagnostic introduced by \cite{2020MNRAS.499.5749J}. In brief, we inspect the data distribution in a 3D space spanned by log([N\,{\sc ii}]/H$\alpha$), log([S\,{\sc ii}]/H$\alpha$), and log([O\,{\sc iii}]/H$\beta$), and fit a demarcation surface that sits between a fiducial SF photoionization model and a fiducial AGN photoionization model, where at roughly 90 per cent of the H$\alpha$ intensity is contributed by SF. The final sample bounded by the demarcation surface is coloured in green and yellow, while the rest of the spaxels are coloured in black and white. 
The projected demarcation lines in the original BPT diagrams are further out compared to the demarcation lines of \cite{kewley2001}, \cite{kauffmann2003}, and \cite{stasinska2006}, but close to the empirical demarcations obtained by using the kinematics of ionized gas \citep{law2021}.

Comparing the three SF models we computed in this work (which are plotted as two dimensional grids spanning a range of metallicities and ionization parameters) to the data, we can see that in the [N\,{\sc ii}]-BPT diagram, the model with $F_* = 0.5$ is closest to the center of the SF locus. Since the depletion factor of nitrogen is almost independent of $F_*$, the change of the model grid along the $x$-axis is purely a result of changing electron temperature. As $F_*$ becomes larger, $\rm T_{\rm e}$ increases and thus the model is moving rightwards. The change along the $y$-axis is only obvious at median-to-high metallicity, which we have already seen in Fig.~\ref{fig:line_strength}. Interestingly, in the [S\,{\sc ii}]-BPT diagram, the trend with $F_*$ is reversed. The depletion in sulfur overwhelms the increase in $\rm T_{\rm e}$. It now appears that the model with zero depletion provides the best-fit. It is noteworthy that in practices of photoionization modelling of H\,{\sc ii} regions, sulfur is usually assumed to have zero or negligible depletion. While in this work, our adopted relation describes a significant depletion of sulfur even when $F_* = 0.5$. Jenkins09 warned that the determination of the depletion of sulfur through observations of stellar absorption might suffer from saturation of sight lines. Therefore, there are still uncertainties associated with the prescription for sulfur.
Regardless, our current set of comparisons show that the observed \hii\ regions in the [N\,{\sc ii}]-BPT diagram prefers the model with $F_* = 0.5$, while the same \hii\ regions in the [S\,{\sc ii}]-BPT diagram prefers the model with $F_* = 0$.

The [O\,{\sc i}]-BPT diagram is the most interesting but also the problematic one. First of all, the extreme starburst line cannot bound a large number of H\,{\sc ii} regions identified using the other two diagrams. In addition, current photoionization models fail to fit this upper envelope of the data as well \citep[see e.g.][]{law2021}.
Despite the change of the model points at low and high metallicities with increasing $F_*$, the upper envelopes of the models remain nearly unmoved. This is indicative of a balance between the increasing depletion of oxygen and the decreasing abundance of oxygen at fixed metallicity near the envelope, which also shows that the depletion factor is not the culprit for the mismatch between the models and data.

\begin{figure*}
\includegraphics[width=\columnwidth]{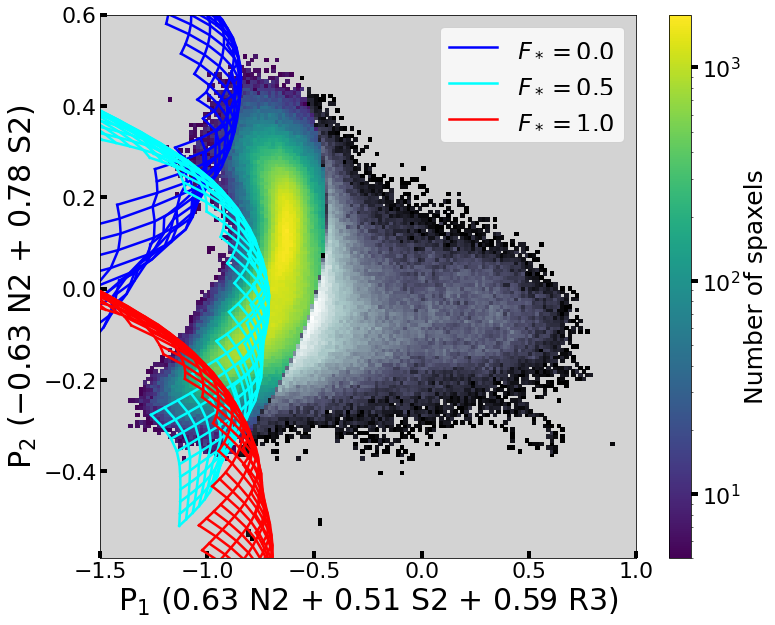}
\includegraphics[width=\columnwidth]{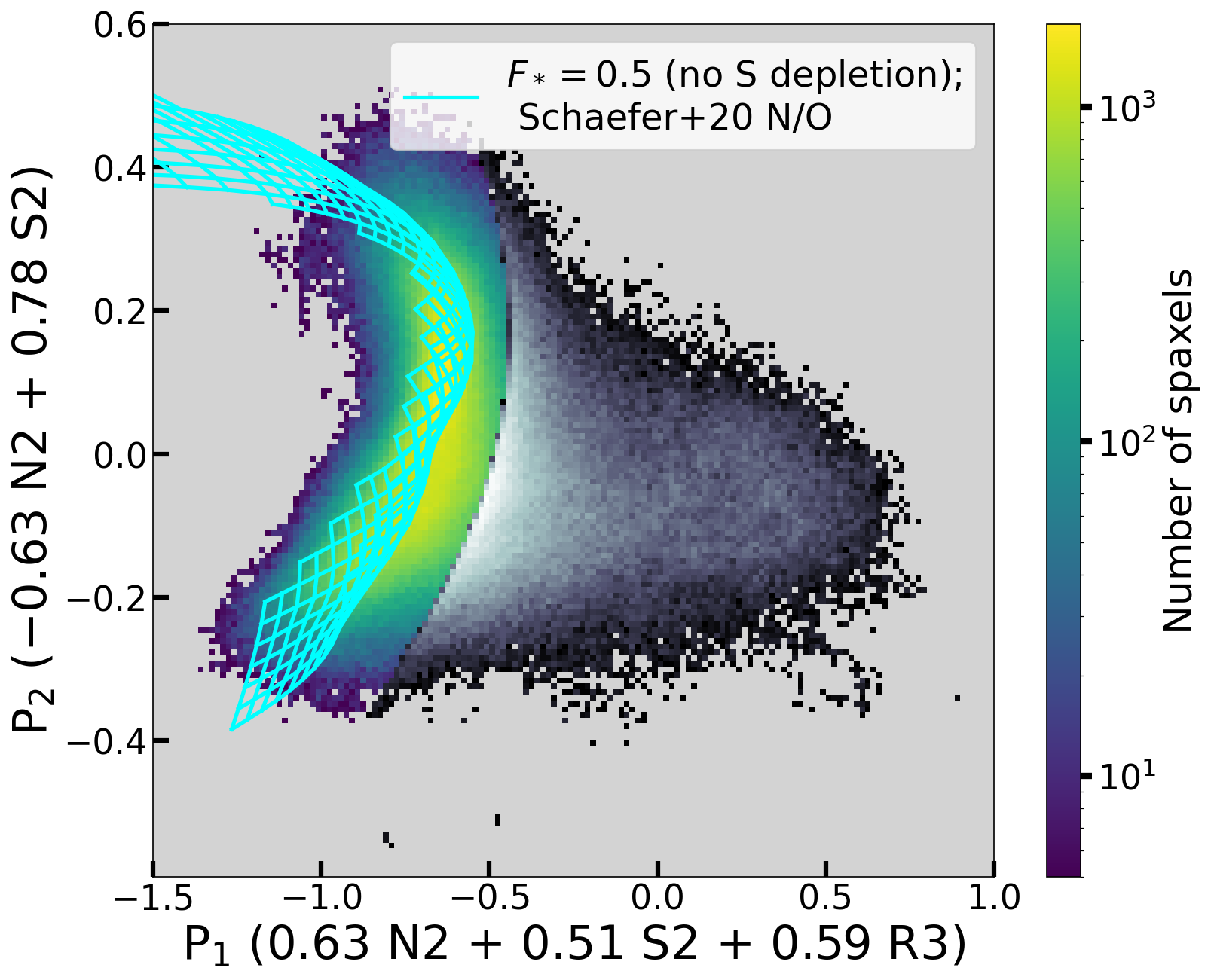}
\caption{Distribution of MaNGA data in a specific projection \citep[$P_1$-$P_2$ projection;][]{2020MNRAS.499.5749J} of a three-dimensional line ratio space. Here N2, S2, and R3 stand for log([N\,{\sc ii}]/H$\alpha$), log([S\,{\sc ii}]/H$\alpha$), and log([O\,{\sc iii}]/H$\beta$), respectively. The two axes $P_1$ and $P_2$ are linear combinations of N2, S2, and R3. 
The projection was constructed in a way that makes the data locus and the model appear nearly edge-on.
The model grids are also projected into this plane but are first interpolated and then truncated so that only the parts of the models that cover the middle 98 per cent of the data along the third dimension perpendicular to both $P_1$ and $P_2$ axes are shown.
\textit{Left:} Model grids with different $F_*$ values. \textit{Right:} A model grid with $F_* = 0.5$ but without depletion of sulfur onto dust grains. In addition, different from the models in the left-hand panel (whose parameters other than $F_*$ are summarized in Table~\ref{tab:model_param}), it adopts an N/O versus O/H relation derived by \protect\cite{schaefer2020}.}
\label{fig:newproj}
\end{figure*}

Fig.~\ref{fig:newproj} compares the models to the data distribution in a reprojected BPT diagram. This diagram uses linear combinations of the logarithms of the BPT line ratios ([N\,{\sc ii}]/H$\alpha$, [S\,{\sc ii}]/H$\alpha$, and [O\,{\sc iii}]/H$\beta$) to place photoionization models on a nearly edge-on view, which puts strong constraints on the locations of the models \citep{2020MNRAS.499.5749J}. One can easily see how well the model grids trace the center of the SF locus. The colour coding of the density distribution of the data is the same as in Fig.~\ref{fig:3bpt}. 
The left-hand panel of Fig.~\ref{fig:newproj} compares models with different values of $F_*$. While the model with $F_* = 0.5$ shows the best consistency with the SF locus, it still shows obvious deviation from the upper part of data locus, which corresponds to \hii\ regions with low metallicities.
As $F_*$ increases, the model grid moves down along the $P_2$ axis but only moves to the right a little at high metallicity. This is because the changes in [N\,{\sc ii}]/H$\alpha$ and [S\,{\sc ii}]/H$\alpha$ have opposite signs and thus add up along $P_2$, while canceling out along $P_1$. In the meanwhile, the change in [O\,{\sc iii}]/H$\beta$ is only significant at high metallicities due to the competing effect of oxygen depletion and overall cooling efficiency. As a result, the model grid exhibits a nearly vertical motion in the plane. We note that our best-fitting value of $F_*$ is consistent with the work by \cite{kewley2019a}, who adopted $\rm \log (Fe_{gas}/Fe_{total}) = -1.5~dex$ for their ISM models. This value is derived from their observations of H\,{\sc ii} regions in the Milky Way and Magellanic Clouds using the Wide Integral Field Spectrograph and corresponds to $F_* = 0.43$.

Based on our choice of input parameters, the model with $F_*=0.5$ still cannot fit \hii\ regions at all metallicities in the MaNGA sample. The uncertainty in the depletion of sulfur might contribute to this offset. If we remove the sulfur depletion entirely from this model but keep the depletion strengths for other elements unchanged, we can improve the model prediction on S2 but worsen the model prediction on N2 (due to the drop in T$_e$). As a result, the model is shifted up in the re-projected diagram, fitting the low-metallicity SF locus slightly better, but fitting the rest of the SF locus worse.
Meanwhile, we note that the sulfur depletion pattern itself, if set as a free parameter, cannot be well determined due to its degeneracy with other nebular parameters.

The choice of the reference abundance set as well as the abundance pattern for the secondary elements could play a role, as it directly influences the overall abundances of different elements. For example, tuning the N/O abundance pattern can compensate the loss in N2 due to the lowered sulfur depletion.
In the right-hand panel of Fig.~\ref{fig:newproj}, we show a model with $F_*=0.5$ but without sulfur depletion. Using an N/O abundance pattern derived by \cite{schaefer2020}, we are able to make this model fit the whole SF locus much better compared to the models in the left-hand panel.
The N/O abundance pattern of \cite{schaefer2020} was derived using the observed [N\,{\sc ii}] \textlambda$ 6583$/[O\,{\sc ii}] \textlambda\textlambda$3726,3729$ in MaNGA \hii\ regions.
The similarity of the model to the model constructed by \cite{2020MNRAS.499.5749J} indicates degeneracy between the depletion pattern and the abundance pattern for N/O. When other model parameters are fixed, as long as the depletion patterns of the major coolants in the optical, in this case N and O, and the input N/O prescriptions prior to depletion produce similar post-depletion abundances for N and O, the resulting relevant line ratios will be similar. Without other independent constraints from observations, it is difficult to break this degeneracy in the model. It is noteworthy that under the depletion scheme we adopted, we need to remove sulfur depletion in the model in order to produce a better fit to the data.

Despite the aforementioned importance of the sulfur depletion, there are few observational constraints on its strength.
Sulfur is often considered as an element of zero depletion when modelling the ISM. However, according to \cite{2009ApJ...693.1236C}, a sulfur depletion of $\sim 0.1$ dex is needed to explain the N/S ratio of the gamma-ray burst (GRB) host galaxies in their sample. On the other hand, the measurement of the local sulfur depletion in different sight lines suffers from small numbers of reliable determinations.
The relation obtained by Jenkins09 using these observations predicts a significant depletion of $\sim 0.4$ dex for sulfur when $F_*$ is only 0.5. Judging from Fig.~\ref{fig:3bpt} and Fig.~\ref{fig:newproj}, it is likely that the depletion of sulfur has been overestimated.
Whereas we also need to be aware of the degeneracy between the sulfur depletion strength and the abundance patterns of other elements.
As an important coolant in the H\,{\sc ii} regions, the abundance of sulfur affects the thermal balance of the clouds and thus is also connected to the strength of other collisionally excited lines. Its depletion pattern might also be related to the long standing problem of the overestimation of [S\,{\sc iii}] \textlambda\textlambda$9069,9531$ by photoionization models \citep[e.g.][]{1989ApJ...345..282G, 2015ApJ...804..100B, mingozzi2020}. To fully settle this issue, we need more direct constraints on the abundance of sulfur as well as other elements in \hii\ regions from observations in the future.

Given the degeneracy between the depletion pattern and the N/O prescription we show, one might wonder whether other input parameters in the photoionization model such as the ionizing spectral energy distribution (SED) could produce degeneracy as well. While changing $F_*$ mainly shifts model along the $P_1$ axis in the $P_1$-$P_2$ diagram, changing the hardness of the ionizing SED mainly shifts the model grid in a nearly perpendicular direction along $P_2$ (see e.g. Figure~7 of \citealp{2020MNRAS.499.5749J}).
Specifically, a harder SED for SF regions enhance both the high-ionization lines and low-ionization lines, thereby shifting the model to the right in this diagram and push it away from the SF locus.
A similar effect can be seen in the original BPT diagrams \citep[e.g.][]{2017ApJ...840...44B,2019ApJ...878....2D}.
On the other hand, fixing other parameters while changing the stellar ionizing SED cannot make the model with the $F_*$ depletion pattern to fit the SF locus significantly better compared to what is shown in the left-hand panel of Fig.~\ref{fig:newproj}.
Therefore, the variation in the ionizing SED is unlikely to produce significant degeneracy with the depletion strength and depletion pattern.

Finally, the dust composition is an interesting ingredient in the model which we have not discussed in detail thus far. We discuss the model assumption in the dust composition in the next section.

\section{Discussions}
The grain physics in {\sc cloudy} has its own separate code (see \cite{2017RMxAA..53..385F} Section 5 and \cite{2006hbic.book.....F} Section 7.9 for further information). {\sc cloudy} incorporates 20 different standard grain types, some of which are graphite and some are silicates. The grain composition used by {\sc cloudy} is one that is consistent with the required $A_{\rm V}/n_{\rm H}$ (an average for extragalactic H\,{\sc ii} regions). A mixture of different grain types provides a composition which helps describe an average of properties of extragalactic H\,{\sc ii} regions. 
Although the grain physics in {\sc cloudy} was encoded separately from applying the Jenkins09 depletion model, this model should still be relevant because it simply lets us how to alter the relative amount of depleted material. We then use these post-depleted abundances in the grain physics calculations. Note currently, the depleted abundances from gas-phase are not necessarily consistent with the abundances in dust-phase. It is difficult to make a fully consistent prescription that takes into account both element depletion and grain composition. However the most realistic and self-consistent a treatment to do so thus far, is by scaling the amount of dust based on the $F_*$ value. This is investigated and discussed in the Paper~\,{\sc i}. The current study utilizes this method for all of the simulations presented. 

Checking if the depleted material is consistent with the material forming the grains is a next step. Since Jenkins09 depletion model was originally developed from observations of neutral clouds, we may first need to build a depletion pattern for H\,{\sc ii} regions similar to the Jenkins09 model -- a discussion on this particular fact is also presented in the Paper~\,{\sc i}.

Finally, note the same analysis as presented in this study can be carried out using fixed nebular metallicity (i.e. fixed post-depletion gas-phase metallicity). To do this with {\sc cloudy} the $F_*$ value in the command line of the input model can be kept constant while scaling the dust according to various values $F_*$. The final result will be equivalent to the ones discussed in this investigation.  

\section{Conclusions}
\label{sec:conclusion}
The primary objective of this study was to explore how varying the collective depletion strength $F_*$, affects the predicted line ratios of a generalized photoionization model. The goal is to help us understand how dust depletion may affect spectral observations as well as provide a constraint on the value of $F_*$ when modelling a collection of H\,{\sc ii} regions.

While it has been long suspected that dust depletion affects the observed line ratios, previous work has not explored the details nor realized the extent of its effects on emission line spectra of H\,{\sc ii} regions. This study has found that spectral line ratios of ionized regions vary non-trivially with the change in depletion strength, especially in R3, S2, and O1. Furthermore, our model predictions indicate that the line ratios are varied by a balance between metallicity, ionization parameter $U$, and depletion factor $F_*$. Metallicity varies the abundance of all elements, while the depletion factor varies fractions of element abundances selectively. 

At low metallicities, temperatures of the gas are high compared to those at high metallicities, as a result of low coolant abundances. 
However, there is little variation in temperature as a function of $F_*$, 
so the trends in emission line ratios are dominated by changes to the depletion strength. 
Likewise, in this regime, the effects of $U$ on the emission line ratios change little with $F_*$ ($F_*$ with S2 and $U$ with R3). Hence the BPT diagrams spanning a smaller range of line ratios than at higher metallicities stipulate that $F_*$ may be constrained independent of $U$ at metallicities lower than the reference set.

At metallicities greater than the reference values, although the electron temperatures are lower, there is significant variation in temperature as a function of depletion strength. In this regime, changes in $F_*$ result in a significant change in the thermal equilibrium of the gas.
As a result of this, collisional excitation rates are increased enhancing the emission line ratios. 
Here, variations in the electron temperature dominate the emission line ratio trends. Unlike at lower metallicities, $F_*$ enhances the trends of the line ratios with $U$ and vice versa. Hence, $U$ must be constrained in order to constrain $F_*$, as the effects on the BPT diagrams by one parameter is dependent on the other.

There are two regimes based on the ionization parameter how varying $F_*$ impacts the size of the ionized region. For $\log(U)\geq-2$, since grains dominantly absorb the incident ionizing photons, the rise in grain abundance with $F_*$ shrinks the ionizing layer. For $\log(U)<-2$, since H atoms dominate in absorbing the ionizing photons, the gas becoming hotter with $F_*$ expands the ionized region. Due to this effect, we conclude that even at low metallicities, $U$ must be constrained in order to constrain $F_*$.

Finally, comparing our photoionization models with the data from the MaNGA survey revealed a preferred depletion factor for this collection of \hii\ regions. The best-fitting value of depletion factor for this particular group of \hii\ regions correspond to $F_*\approx0.5$ as revealed by the [N\,{\sc ii}]- 
and the re-projected BPT diagrams. However, our photoionization models in this study still fail to fit the observed N2, S2, and O1 simultaneously, especially at low metallicities.
Removing the sulfur depletion and adopting a different N/O pattern make the model with $F_*=0.5$ fit the data locus better, which also reveals the degeneracy between the depletion pattern and the N/O pattern.
To solve this issue, we need direct constraints on the depletion of sulfur and a deeper understanding of the photoionization modelling of sulfur and oxygen lines in future works.
Regardless, even though our models do not provide a perfect fit to the data, we have seen that it is still possible to put qualitative constraints $F_*$ by selecting the model which shows the best consistency with the data locus. 

The above paragraphs summarize the main conclusions borne from this study. Some additional caveats must be included to provide a complete picture: 
\begin{enumerate}
    \item Sulfur depletion: Since this depletion pattern was developed for H\,{\sc i} regions, the depletion pattern of sulfur may not be an accurate estimation for H\,{\sc ii} regions. However, sulfur being an important coolant affects the thermal balance of the gas and consequently the emission line ratios of other ions.
    \item At low values of $F_*$, some depletion scale factors in the pattern described by Jenkins09 attain values above unity. However, it is not physical for the abundance of any element to increase from its initial value. The present study avoids this unphysical issue by defaulting the depletion scale factors to 1, if the calculation according to Jenkins09 rises above unity at $F_*=0$.  
    For sulfur, the overestimated abundance could originate from the contamination from \hii\ regions along the sight lines used to measure the sulfur abundances in the H\,{\sc i} gas.
\end{enumerate}

\section*{Acknowledgements}
We thank all the people that have made this paper possible.
CMG was supported by STScI (HST-AR-15018 and HST-GO-16196.003-A).
MC acknowledges support by STScI (HST-AR14556.001-A), NSF (1910687), and NASA (19-ATP19-0188).
GJF acknowledges support by NSF (1816537, 1910687), NASA (ATP 17-ATP17-0141, 19-ATP19-0188), and STScI (HST-AR- 15018 and HST-GO-16196.003-A).
RY acknowledges support by the Hong Kong Global STEM Scholar scheme, by the Hong Kong Jockey Club through the JC STEM Lab of Astronomical Instrumentation program, by the Direct Grant of CUHK Faculty of Science, and by the Research Grant Council of the Hong Kong Special Administrative Region, China (Projects No. 14302522).

\textit{Software:} Python 3.8 \citep{python3}, {\sc cloudy} \citep{2017RMxAA..53..385F}.

\section*{Data Availability}
The data underlying this article are available in the article and the {\sc cloudy} distribution. 
The observational data used by this work comes from the DR15 of SDSS, which is publicly available\footnote{www.sdss.org/dr15/data\_access/}.



\bibliographystyle{mnras}
\bibliography{GrainsAndEmissionLines} 





\bsp	
\label{lastpage}
\end{document}